\documentclass[fleqn,usenatbib]{mnras}

\usepackage{color}
\usepackage[english]{babel}
\usepackage{float}
\usepackage{verbatim}
\usepackage{hyperref}
\hypersetup{linkcolor=red,citecolor=blue,filecolor=cyan,urlcolor=magenta}
\usepackage{newtxtext,newtxmath}
\usepackage[T1]{fontenc}
\usepackage{graphicx}	
\usepackage{amsmath}	
\usepackage{longtable}
\usepackage{deluxetable}
\usepackage{lineno}

\title{Puzzling High-Velocity Calcium Absorption Features Of Type Ia Supernovae}

\author[X. Zhao]{Xulin Zhao$^{1,2}$
\thanks{Contact e-mail: zhaoxulin@139.com}
	\\
	$^{1}$Tianjin Key Laboratory of Quantum Optics and Intelligent Photonics, School of Science, Tianjin University of Technology, Tianjin 300384, China
	\\
    $^{2}$School of Science, Tianjin University of Technology, Tianjin 300384, China
}

\begin{document}

\label{firstpage}
\pagerange{\pageref{firstpage}--\pageref{lastpage}}
\maketitle

\begin{abstract}
	Absorption features Ca II NIR and Ca II H\&K of type Ia supernovae (SNe Ia) are characterized by their strong high-velocity features (HVFs). We find that, for these two features of calcium there is a puzzling anti-correlation between the line strengths of HVF and photospheric (PHO) components, and an unexpected positive correlation between the velocity difference and line strength ratio of HVF and PHO components. In comparison, HVFs of Si II $\lambda$6355 and O I $\lambda$7773 show a positive correlation between the line strengths of HVF and PHO components, and no clear correlation between the velocity difference and line strength ratio of the two components. The differences may be associated with the fact that calcium was mostly synthesized in deeper layers than silicon and oxygen, and thus experienced much more serious blocking by substances in outer layers. These observations can shed light on the physics of HVFs.
\end{abstract}


\section{Introduction}

Type Ia supernovae (SNe Ia) have been precise distance indicators because of their relatively homogeneous brightness \citep[e.g.][]{Phillips93,Phillips99, Riess98, Riess19, Perlmutter99, Freedman19,Jha19}. They are believed to be caused by explosive thermonuclear burning triggered either in the core of a nearly-Chandrasekhar-mass white dwarf (WD) \citep[e.g.][]{Nomoto82,Khokhlov91,Hillebrandt00,Maeda10,Maeda18, Flors20} or on the
surface of a sub-Chandrasekhar-mass WD \citep[e.g.][]{Kushnir13, Pakmor13,Blondin18, Shen18, Flors20}. There are two popular scenarios proposed for their progenitor systems: (1) single degenerate scenario (SD) consists of a CO white dwarf (WD) and a non-degenerate companion star such as a main-sequence or red-giant star \citep[e.g.][]{Whelan73, Nomoto82}. An example is PTF 11kx which shows a low velocity ($<1,000$ km/s) Ca II H\&K along with narrow He I $\lambda$5876 in its spectra \citep[e.g.][]{Dilday12}. The low velocity Ca substances were likely ejected from a symbiotic nova progenitor, while He lines indicate a non-degenerate companion star. (2) double degenerate scenario (DD) consists of two WDs \citep[e.g.][]{Iben84, Webbink84}. The DD scenario may explain, for example, the lack a companion star with luminosity greater than a few percent of the sun associated with the famous nearby SN 2011fe \citep[e.g.][]{Li11}. 

Detailed information about SN Ia, for example the element distributions, the kinematics or the temperature are mostly obtained through spectroscopic analysis of absorption features. Important features include at least: Si II $\lambda$6355 whose velocity is often chosen to represent the velocity of the ejecta; C II $\lambda$6580 which serves as an indicator of unburnt carbon \citep[e.g.][]{Thomas11,Folatelli12}; Si II $\lambda$5972 as an indicator of the photospheric temperature and peak luminosity as well \citep{Nugent95,Blondin12,Zhao15}; variable sodium lines, H and He lines as indicators of thick CSM \citep[e.g.][]{Patat07,Blondin09,Dilday12}. According to the velocity gradient of Si II $\lambda$6355, SNe Ia can be grouped into subgroups of `Faint', `high-velocity gradient' (HVG), and `low-velocity gradient' (LVG) SNe Ia \citep{Benetti05}. Based on the line strengths of Si II  $\lambda$6355 and  $\lambda$5972, SNe Ia are grouped into `core-normal' (CN), `broad-line' (BL), `cool', and `shallow silicon' (SS) SNe Ia \citep{Branch09}. According to the velocity of Si II $\lambda$6355 ($V_{Si6355}$), SNe Ia are mostly grouped into `Normal-Velocity' (NV) SNe Ia which have $V_{Si6355} \lessapprox 12,000$ km/s and `High-Velocity' (HV) SNe Ia which have $V_{Si6355} > 12,000$ km/s \citep{Wang09,Wang13}.

The formation of the absorption features on ejecta of SN Ia mostly occur by coherent scattering \citep[e.g.][]{Bongard08}, with a photon energy lower than ionization energy but high enough to excite the electron temperately to the higher level ($E_{higher}$). The `vibrating' electron soon fell back to the lower level ($E_{lower}$), and emitted a photon with the same energy in random direction. Generally speaking, the line strength of the feature increases with the oscillation strength and the ion abundance, and decrease with the lower level ($E_{lower}$)\footnote{ref: \href{https://supernova.lbl.gov/~dnkasen/tutorial/}{https://supernova.lbl.gov/$\sim$dnkasen/tutorial/}} and excitation energy $E_{exc}=E_{higher}-E_{lower}$ \citep[e.g.][]{Zhao21}. The features are mainly produced by absorption in photosphere (PHO), but they are also contributed by a high-velocity feature (HVF) component, which is attributed to absorption in distant regions. The velocity of a HVF is typically between 17,000 and 25,000 km/s. The most studied one is the HVF of Ca II NIR. Its correlations with observables such as the decline rate, the B-V color and the host galaxies have been investigated with different samples \citep[e.g.,][]{Childress14,Maguire14,Silverman15}, in an attempt to improve the calibration of the peak luminosity. It has been suggested that HVFs may be related with enhancements of element abundance, matter density, or ionization in outermost layers of the ejecta \citep[e.g.][]{Gerardy04,Mazzali05,Tanaka08}. 

Observations suggest that the HVFs are closely correlated with their corresponding PHO components \citep{Zhao15,Zhao16}. This means that the substances of HVFs might be originally located in inner or near surface layers of the progenitor, and then moved to the outermost layers with their high velocities. A critical clue is an anti-correlation between the HVFs of Si II $\lambda$6355 and O I $\lambda$7773 for NV SNe Ia, which may be associated with burning of oxygen to silicon on the surface of the progenitor. These observations place new constraints on explosion
models \citep[e.g.][]{Kato18}. Following our previous works \citep{Zhao16}, we carried out an investigation of the HVF of Ca II H\&K. Compared with Ca II NIR triplet whose main contributing lines (i.e. Ca II $\lambda$s 8542 and 8662) are quite separated in wavelength, Ca II H\&K 's two contributing lines at 3934 and 3968 \AA~ are much closer in wavelength. This significantly reduces the blending between the HVF of redder contributing line and the PHO of bluer contributing line, and hence improve the accuracy of the measurement. Note that Ca II H\&K may be affected by Si II $\lambda$3850, but in most case very insignificantly \citep[similar conclusions were made in][]{Childress14, Maguire14, Silverman15}. For example, in the spectra we show latter, no clear sign of Si II $\lambda$3858 is seen. This is consistent with theoretical expectation. Compared with Si II $\lambda$4130 which itself is very weak (mean pseudo-equivalent width (pEW) at -10 days $\approx$ 13 Å for NV SNe Ia, no HVF), Si II $\lambda$3850 has a much smaller oscillation strength (0.5 vs. 5.1), and thus should be even weaker and has no HVF. But if necessary, this line can be removed by using a multiple Gaussian fitting.

This paper is organized as follows. In Section \ref{Sect2} we briefly describe the sample and measurement procedure. In Section \ref{Sect3} we focuses on the behaviors of the HVFs of Ca lines. Summary and discussion about the origin of HVFs of Ca are given in Section \ref{Sect4}.

\section{Data and Measurement}
\label{Sect2}

Measurement results for lines Si II $\lambda$6355, O I $\lambda$7773 and Ca II NIR have been reported in our previous works \citep{Zhao15, Zhao16}\footnote{Correction for Ca II NIR of SN 2013gs at -7.5 days: $V_{PHO}$= 13,200 km/s, $pEW_{PHO}$ = 21 \AA, $V_{HVF}$= 19,742 km/s, $pEW_{PHO}$ = 9 \AA.}. Measurement results for line Ca II H\&K are listed in Table \ref{Tab1}. The spectra are mainly from the database of the Harvard-Smithsonian Center for Astrophysics (CfA) Supernova Program \citep{Matheson08, Blondin12}. This allows us to restrict the diversity that is introduced by using data from different sources, as the observation conditions and spectral reductions could be very different for different observers. Others are mostly from the Berkeley Supernova Program \citep{Silverman12a, Silverman12b, Stahl20} and the Carnegie Supernova Project \citep[CSP,][]{Folatelli13}.

As usual, absorption features on the spectra (divided by the pseudo-continuum) are fitted with a multiple Gaussian function. For example, Ca II H\&K is fitted with a double Gaussian function, which follows as $f(\lambda)=c_1exp(\frac{-(\lambda_1-\lambda_0)^2}{2\sigma_1^2}) +c_2exp(\frac{-(\lambda_2-\lambda_0)^2}{2\sigma_2^2})$, where `$c$' represents the amplitude, `$\sigma$' represents the dispersion, $\lambda$ represents the wavelength, and $\lambda_0$ represents the rest-frame wavelength. Line velocity is calculated by $V=(\lambda^2-\lambda_0^2)/(\lambda^2+\lambda_0^2)\times 3\times 10^5$(km/s), while the pseudo-equivalent width $pEW=\int_{\lambda_1}^{\lambda_2} \frac{F_C(\lambda)-F(\lambda)}{F_C(\lambda)} d\lambda$, where $F$ represents the flux density and $F_{c}$ represents the continuum flux density. 

Rest-frame wavelength of Ca II NIR is as usual set to be 8567 \AA. But, it could be underestimated. A simple estimation weighted by oscillator strength would suggest a rest-wavelength of about 8584 \AA. For Ca II H\&K, the rest-wavelength is set to be 3945 \AA. Measurement uncertainties are mainly due to continuum fitting (continuum is often more irregular at early phases than near maximum light), line blending (including the blending between PHO and HVF components), flux error (relatively higher at early phases than near maximum light, and relatively higher in violet and near-infrared bands than central optical band) and atmospheric absorption. 

\section{HVFs of calcium}
\label{Sect3}
 
In this section, we report our measurement results of the HVFs of lines Ca II NIR and Ca II H\&K. Of most interest are some that are different with the HVFs of Si II $\lambda$6355 and O I $\lambda$7773. 

\subsection{Comparison between Ca II NIR and Ca II H$\&$K}

\begin{figure*}
	\includegraphics[width=1.8\columnwidth]{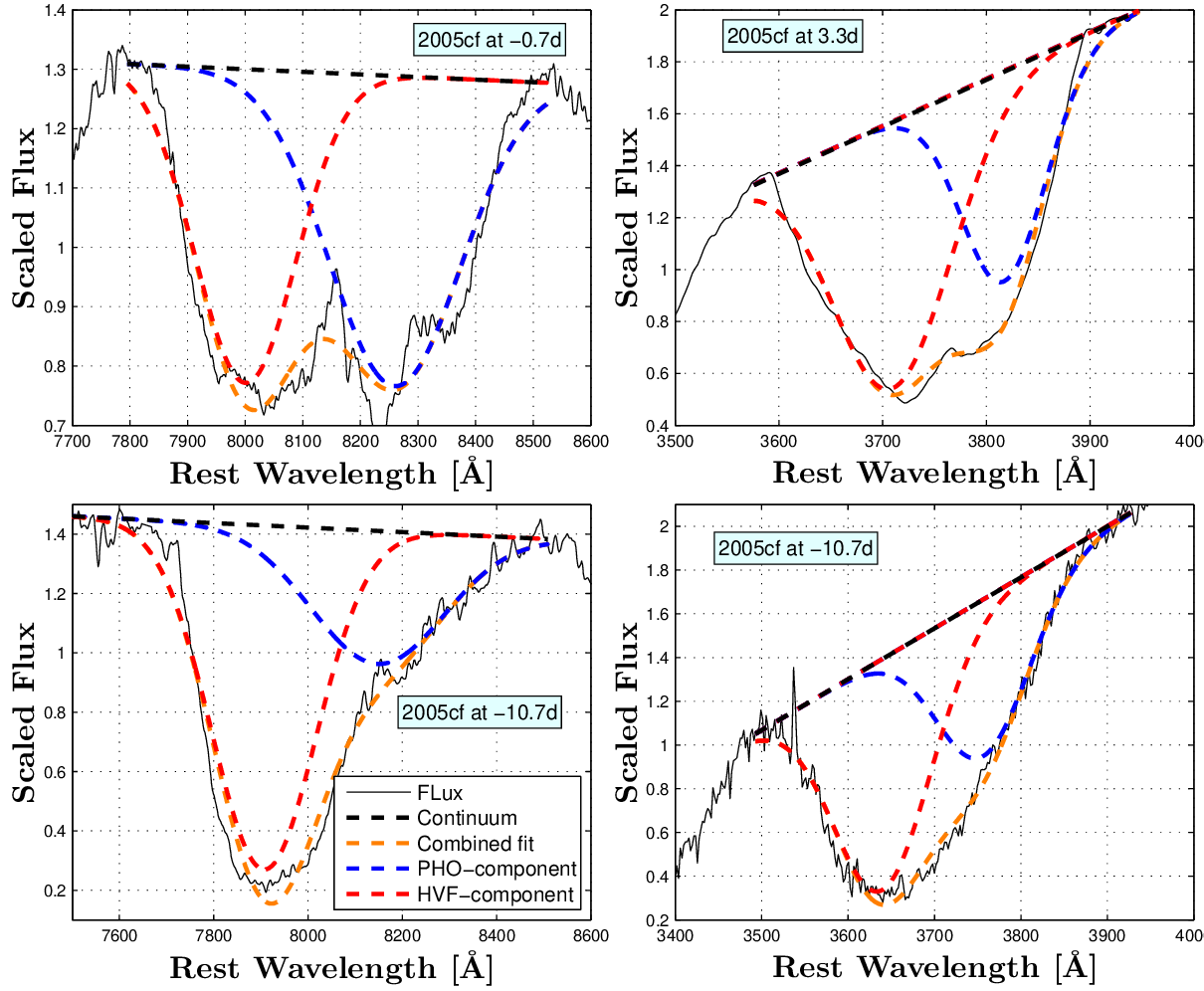}
	\caption{Gaussian fittings for Ca II NIR and Ca II H\&K of SN 2005cf at different phases. Upper-left panel: at $t=$-0.7 days; Upper-right panel: at $t=$3.3 days; Lower panels: at $t=$-10.7 days.}
	\label{Fig1} 
\end{figure*}

\begin{figure*}
	\includegraphics[width=1.8\columnwidth]{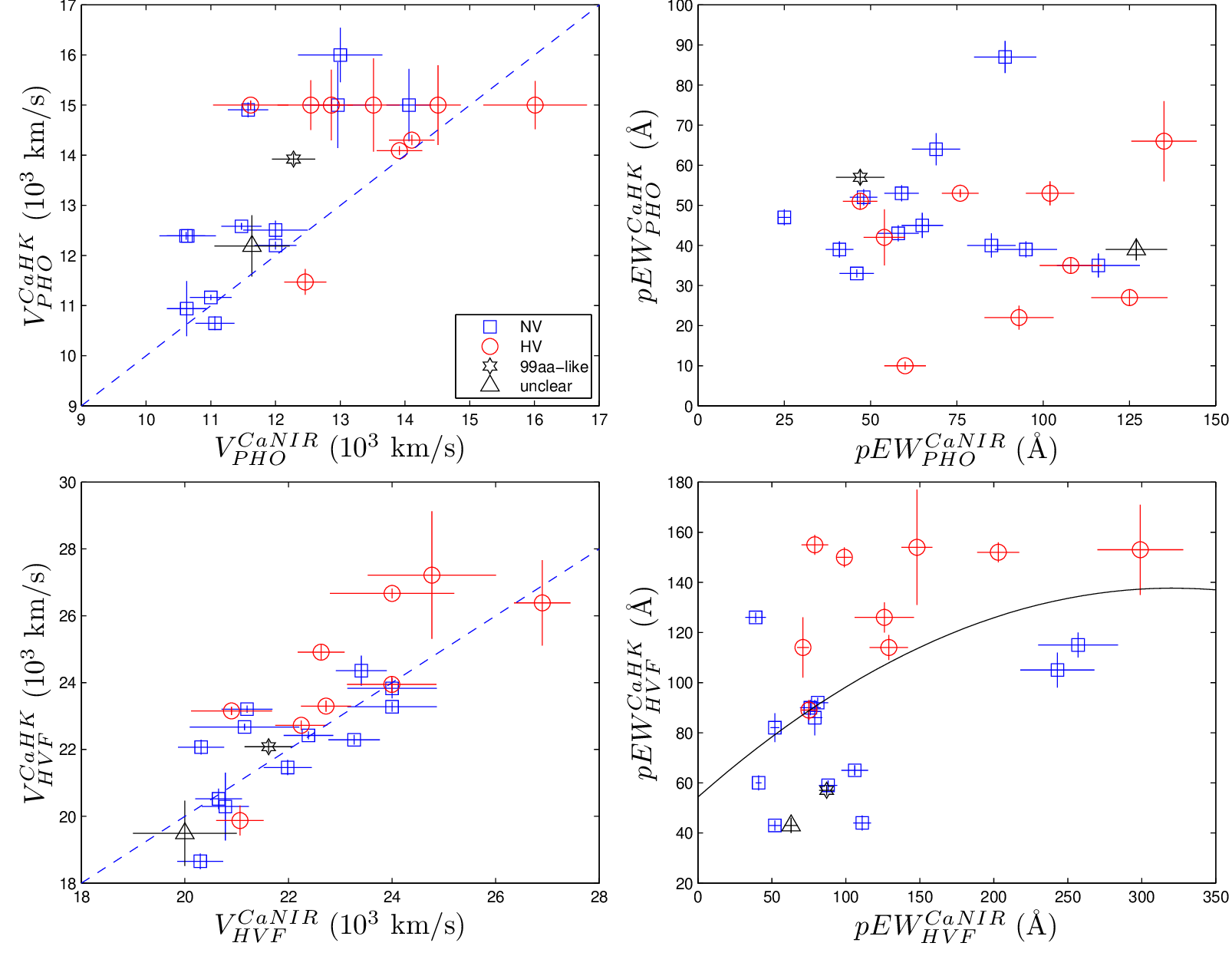}
	\caption{\label{Fig2} A comparison between the velocities ($V$) and line strengths ($pEW$) of Ca II H\&K and Ca II NIR. The phase is restricted to $t=-10\pm2.5$ days. The sample has been split into subclasses as defined by \citet{Wang09}. Photospheric velocity was assumed to be less than 15,000 km/s for most objects, and 16,000 km/s for a few HV SNe. Note that the sample size is seriously reduced for requiring the spectrum to cover both Ca H\&K and Ca II NIR ($3,400 \sim 8,700$ \AA). Pearson's linear correlation coefficient is 0.67, -0.03, 0.82 and 0.48 for the whole sample presented in the upper-left, upper-right, lower-left and lower-right panel, respectively. Spearman's rank coefficients of these correlations are 0.7, -0.11, 0.86 and 0.45. Grading standards: correlation coefficient ranging 0$\sim$0.2, 0.2$\sim$0.4, 0.4$\sim$0.6, 0.6$\sim$0.8 and 0.8$\sim$1.0 indicates, respectively, a very weak, weak, moderate, strong and very strong correlation. Dash lines are shown to guide the eyes, while the solid line in the lower right panel is a second-order polynomial fit to the data. Dash lines in the left panels are drawn to guide the eyes (with scales of 1:1), while the solid line in the bottom-right panel is a second-order polynomial fit to the data.}
\end{figure*}

Examples of our Gaussian fitting are shown in Fig.\ref{Fig1}. Unlike HVFs of Si II $\lambda$6355 and O I $\lambda$7773 which vanish by about -6 days, HVFs of calcium could remain prominent near maximum light. At early phases, the HVFs of calcium could be highly saturated. For example, in the lower panels of Fig.\ref{Fig1}, the saturation degree is $1-0.2/1.4 \approx 86\%$ and $1-0.3/1.4 \approx 79\%$, respectively, for HVFs of Ca II NIR and Ca II H\&K of SN 2005cf at -10.7 days. Saturation may be one reason that HVF of Ca II H\&K appears to have a ceiling pEW around 160 \AA~(Fig.\ref{Fig2}, lower-right panel).

The strong line strengths of the HVFs of Ca II NIR and Ca II H\&K are believed to be related with its very low $E_{lower}$ and $E_{exc}$. To see how significantly the two factors affect, we compare the velocities and line strengths (quantified by the `pEW') of Ca II NIR and Ca II H\&K in Fig.\ref{Fig2}. Pearson’s linear correlation coefficients and Spearman’s rank coefficients are presented in the caption of the figure, mostly confirming the expectation of positive correlations between velocities or line strengths of Ca II NIR and Ca II H\&K. The only exception is the correlation of photospheric line strengths, with Pearson and Spearman's coefficients being -0.03 and -0.11 respectively. The correlation between line strengths of the HVFs of Ca II NIR and Ca II H\&K is also relatively weak as compared to the correlations of velocities, with Pearson and Spearman’s coefficients being 0.48 and 0.45 respectively.

As one can see from the upper-right panel of Fig.\ref{Fig2}, line strengths of the photospheric component of Ca II H\&K are mostly between 30 and 60 \AA~(centred around 45 \AA), while that of Ca II NIR are mostly between 30 and 120 \AA~(centred around 77 \AA). And, as one can see from the lower-right panel of Fig.\ref{Fig2}, line strengths of the HVF component of Ca II H\&K are between 40 and 160 \AA~(centred around 99 \AA), while those of Ca II NIR are mostly between 40 and 300 \AA~(centred around 113 \AA). Therefore the line strength of Ca II NIR is at least comparable to that of Ca II H\&K. Considering that Ca II H\&K has a much larger oscillator strength (2.0 vs. 0.7) and lower $E_{lower}$ (0.1 eV vs. 1.7 eV) than Ca II NIR, this result may be suggesting a significant effect of the $E_{exc}$ (3.1 eV vs. 1.4 eV) for the two lines. 

On the other hand, the effects of $E_{lower}$ and $E_{exc}$ seem to be less important to the velocities of HVFs. As shown in the lower left panel, Ca II NIR and Ca II H\&K have similar velocities. The small difference ($\Delta v\approx$ 500 km/s) could be due to underestimation of the rest-wavelength of Ca II NIR as mentioned in section \ref{Sect2}, i.e., $(8584-8567)/8584\times3\times10^5\approx600$~km/s. Similarly, the HVFs of Si II $\lambda$6355 and O I $\lambda$7773 have very similar velocities \citep{Zhao16}, though they have different $E_{lower}$ and $E_{exc}$. A possible explanation is that HVF layers might be much thinner, and so has a more uniform velocity. 
 
\subsection{Anti-correlation between the Line Strengths of PHO and HVF components}\label{Sect3.2}

\begin{figure*}
	\includegraphics[width=1.8\columnwidth]{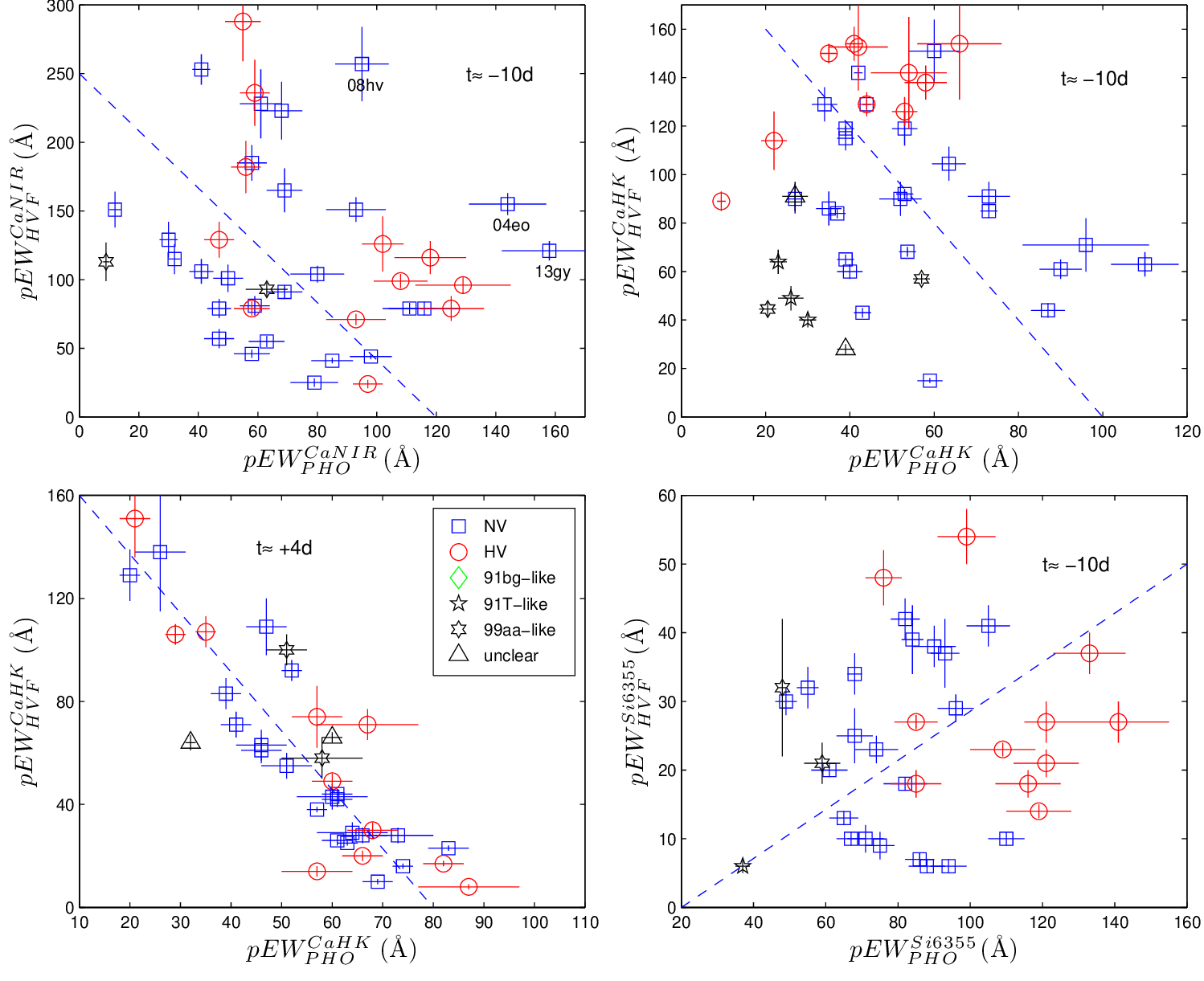}
	\caption{\label{Fig3} Correlation between the line strengths ($pEW$) of PHO and HVF components. Upper-left panel: for Ca II NIR at $t=-10 \pm 2.5$ days; Upper-right panel: for Ca II H\&K at $t=-10 \pm 2.5$ days; Lower-left panel: for Ca H\&K at $t=+4 \pm 1$ days; Lower-right panel: for Si II $\lambda$6355 at $t=-10 \pm 2.5$ days (objects with $pEW_{HVF}^{Si6355}<5$\AA~ are not included, as their HVFs are too weak to distinguish from the broad line). Pearson's linear correlation coefficient is -0.18, -0.08, -0.77 and 0.22 for the whole sample presented in the upper-left, upper-right, lower-left and lower-right panel, respectively. Spearman's rank coefficients of these correlations are -0.22, 0.06, -0.78 and 0.16. Dash lines are drawn to guide the eyes, with scales of roughly 1:2 for top and bottom-left panels, and 3:1 for bottom-right panel.}
\end{figure*}

As mentioned before, the correlations between HVF and PHO components may provide important clues to the origin of HVFs. Figs.\ref{Fig3} shows the correlations for Ca II NIR and Ca II H\&K. It appears that for Ca II NIR and Ca II H\&K the line strengths of the HVF and PHO components are anti correlated. The Pearson's correlation coefficient for Ca II NIR at $t=-10 \pm$ 2.5 days is p$\approx$-0.31 for NV subclass excluding outliers SNs 2004eo, 2008hv, 2013dy, while the Spearman’s rank coefficient is -0.37. It is not clear why the three outliers behave differently, but the correlation for NV subclass is too weak to resist their influences (Pearson and Spearman coefficients reduced to -0.07 and -0.13). It is also better to use the whole sample rather than NV subclass (which is believed to be more homogeneous) to increase universality, but the correlation is much weaker for the whole sample (p= -0.18, Spearman coefficient = -0.22). For Ca II H\&K at $t=-10 \pm$ 2.5 days, p= -0.35 (including all NV SNe), while the Spearman’s rank coefficient is -0.28. The correlation becomes tighter near maximum light. An example is shown in the lower left panel of Fig.\ref{Fig3}. For Ca II H\&K at $t=+4 \pm$ 1 days, the Pearson's coefficient is p= -0.77 (whole sample), and, for Ca II H\&K at $t=0 \pm$ 1 days, p= -0.64 (whole sample). Lower degree of saturation, better sampling (e.g., phase rang = $\pm$ 1 vs. $\pm$ 2.5 days) and smaller uncertainty (e.g., less blending between HVF and PHO components) might be among the reasons that the correlation grows tighter with time. Also noticed is that, the slope $\Delta (pEW_{HVF}^{Ca})/\Delta (pEW_{PHO}^{Ca}) \approx 2$ appears to be the same for Ca II NIR and Ca II H\&K, and the same at -10 and +4 days. 

This anti-correlation may reflect the conservation law of mass, i.e. the more calcium escaped to the outermost layer (becoming HVF of Ca), the less calcium would be left in the photosphere (or even deeper layers that can not be seen in the spectrum, see Section \ref{Sect4} for other possible explanations). For Si II $\lambda$6355, however, as shown in the lower-right panel of Fig.\ref{Fig3}, the correlation is positive: $pEW_{HVF}^{Si6355}\approx pEW_{PHO}^{Si6355}/3$. This correlation is relatively weak, with Pearson and Spearman’s coefficients being 0.22 and 0.16 respectively. A tighter correlation for O I $\lambda$7773 can be found in \citet{Zhao16}, with a similar slope: $pEW_{HVF}^{O7773}\approx pEW_{PHO}^{O7773}/3$.

\subsection{Correlation between velocity difference $\Delta V_{HP}$ and line strength ratio $R_{HP}$}

\begin{figure*}
	\includegraphics[width=1.8\columnwidth]{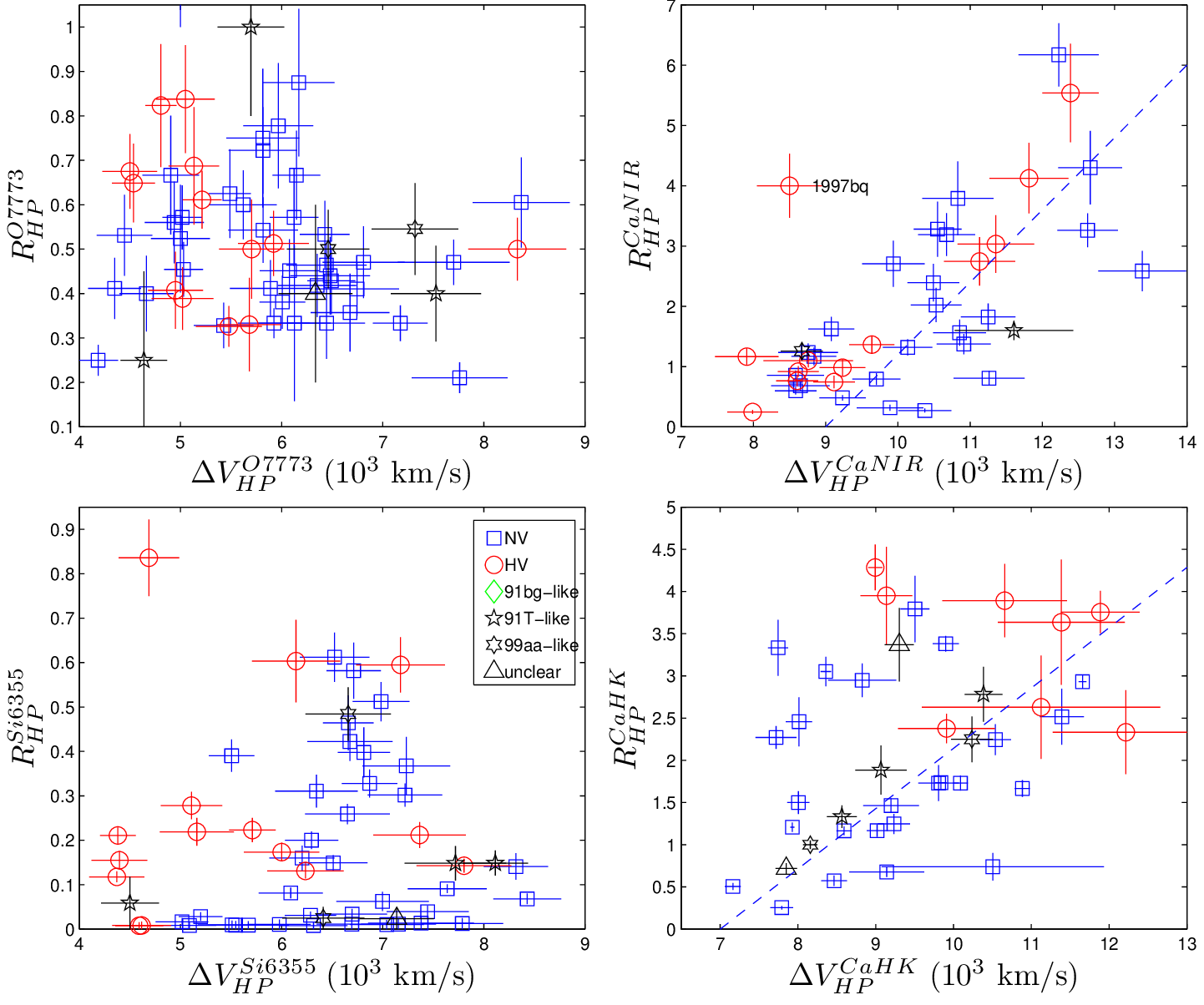}
	\caption{\label{Fig4} Correlation between velocity difference ($\Delta V_{HP}=V_{HVF}-V_{PHO}$) and line strength ratio ($R_{HP}=pEW_{HVF}/pEW_{PHO}$) for SNe Ia at phase $t=-10\pm 2.5$ days. Upper left panel: for line O I $\lambda$7773; Upper right panel: for line Ca II NIR; Lower left panel: for line Si II $\lambda$635; Lower right panel: for line Ca II H\&K. Pearson's linear correlation coefficient is -0.16, 0.65, 0.03 and 0.41 for the whole sample in the upper-left, upper-right, lower-left and lower-right panel, respectively. Spearman's rank coefficients of these correlations are -0.16, 0.64, 0.12 and 0.41. Dash lines in the right panels are drawn to guide the eyes, with scales of roughly 1:1 for top-right panel, and 3:2 for bottom-right panel.} 
\end{figure*}

To further explore the correlations between HVF and PHO components, we investigate the correlation between their velocity difference $\Delta V_{HP}=V_{HVF}-V_{PHO}$ and their line strength ratio $R_{HP}=pEW_{HVF}/pEW_{PHO}$. In theory, a larger velocity difference would lead to a larger distance between the HVF and PHO layers, and hence reduce the absorption efficiency (of the photons emitted from photosphere) in HVF layer. Therefore an anti-correlation was expected. But on the contrary, as shown in the right panels of Fig.\ref{Fig4}, for the two features of calcium, $\Delta V_{HP}$ is positively correlated with $R_{HP}$. Pearson's coefficient is p$\approx$0.67 (whole sample) or 0.73 (excluding SN 1997bq) for Ca II NIR, and, p$\approx$0.41 for Ca II H\&K (whole sample). A possible explanation is that a greater $\Delta V_{HP}$ means a greater kinetic energy of the calcium, and so a greater chance to escape to the outermost layer (see Section \ref{Sect4} for more detailed discussions and other possible explanations).

For Si II $\lambda$6355 and O I $\lambda$7773, however, the correlation is very weak or non-related. A possible explanation is that as much lighter elements, the silicon and oxygen were located near the surface. This means much less substances blocking the way to the outermost layer. Pearson's correlation coefficient p= 0.03 and -0.16 for Si II $\lambda$6355 and O I $\lambda$7773, respectively.

\section{DISCUSSION AND SUMMARY}
\label{Sect4}

Our measurement results show that the HVFs of Ca II behave quite differently with HVFs of Si II and O I. They are much stronger and more persistent, and correlate very differently with their corresponding PHO components. Below we summarize our major findings, and discuss on their possible reasons.

(1) For lines Si II $\lambda$6355 and O I $\lambda$7773, the pEWs of HVF and PHO components are positively correlated, i.e. $pEW_{HVF} \approx pEW_{PHO}/3$ (a relatively weak correlation for line Si II $\lambda$6355 is presented in Section \ref{Sect3.2}, while a relatively strong correlation for line O I $\lambda$7773 is presented in \citet{Zhao16}). This correlation may be caused by radiation conditions that changed in the same direction in PHO and HVF layers, such as the temperature and luminosity. For example, 1991T-like objects are believed to have a relatively higher temperature in both photosphere and HVF layer, while dimmer objects may have a colder photosphere and HVF layer.

(2) For lines Ca II NIR and Ca II H\&K, however, the pEWs of HVF and PHO components are anti correlated, i.e. $pEW_{HVF} \approx -2\times pEW_{PHO} + c$, where $c$ is a constant (the correlation is relatively weak at early phases, and relatively strong near maximum light). Below are three possible reasons:

(I) The first possible reason is line competition. The `pEW' reflects the absorption efficiency $N_{absorbed}/N_{total}$ (`N' represents the number of photons), which depends on ion abundance. When the ion was insufficient, there will be competition among all features of the same ion. However, since the HVF and PHO layers were two separate, independent layers with large distance (due to the huge velocity difference), the competition was unlikely the main reason. 

(II) The second possible reason is the effect of $E_{lower}$. Both Ca II H\&K and Ca II NIR have a rather low $E_{lower}$, but explaining with it may be difficult: (a) Firstly, the effect depends on multiple factors, most of which change in the same direction in HVF and PHO layers, for example the temperature. (b) Secondly, as mentioned before, the two layers were basically independent. (c) Thirdly, the slop $\Delta (pEW_{HVF}^{Ca})/\Delta (pEW_{PHO}^{Ca}$) is almost the same ($\approx$ -2) for Ca II NIR and Ca II H$\&$K which have different $E_{lower}$, and, is almost the same at -10 and +4 days when the ionization conditions changed.

(III) The third possible reason: similar to the anti-correlation between HVFs of Si and O \citep{Zhao16}, this anti-correlation may also reflect the law of mass conservation: Suppose most calcium was synthesized in inner layers, part of the substances might successfully escape to the outermost layer and form the HVF, while others were blocked in the photosphere or even deeper layers. The more calcium escaped, the less would be left in inner layers. For Si and O, the blocking might be much weaker as the elements were synthesized/located near the surface, explaining the fact that there is no anti-correlation between pEWs of HVF and PHO components for the Si and O features.

There are other clues supporting this explanation: (a) As shown in \citet{Zhao16}, the photospheric velocities of Ca II NIR, O I $\lambda$7773 and Si II $\lambda$6355 are quite similar ($t\approx$-10 days), suggesting possibly serious collisions and mixing near the surface. (b) HVFs of Ca have much larger velocity than other HVFs ($\Delta V \approx$ 4,300 km/s), this contrasts with the fact that HVFs of O and Si have similar velocities. (c) A constant-like slop $\Delta (pEW_{HVF}^{Ca})/\Delta (pEW_{PHO}^{Ca}$) for both Ca II NIR and Ca II H\&K.

(3) For lines Ca II NIR and Ca II H\&K, there is a strong positive correlation between $\Delta V_{HP}$ and $R_{HP}$. A possible explanation is that some variable indirectly connects $\Delta V_{HP}$ and $R_{HP}$ and forms the positive correlation. But such intermediate variable has not been found (e.g., for Ca II H\&K in Fig.\ref{Fig4}, Pearson's coefficient p($\Delta m_{15}$,$\Delta V_{HP}$)= -0.17 while p($\Delta m_{15}$,$R_{HP}$)=+0.21). Also, it might be difficult to explain why the intermediate variable only correlates strongly with Ca II features.

Another possible explanation is that the correlation reflects the truth that a greater $\Delta V_{HP}$ corresponds to a greater kinetic energy of the HVF substances, and thus a greater chance to escape the blocking. For Si and O, however, the correlation is very weak or non-related, which can be explained as the `blocking effect' was insignificant for Si and O that were originally located near the surface. 

In conclusion, our measurements reveal two important correlations for Ca II NIR and Ca II H\&K: an anti-correlation between pEWs of the PHO and HVF components, and a positive correlation between $\Delta V_{HP}$ and $R_{HP}$. In comparison, for O I $\lambda$7773 and Si II $\lambda$6355, $pEW_{HVF} \approx pEW_{PHO}/3$, while correlation between $\Delta V_{HP}$ and $R_{HP}$ is too weak to identify. The results are more consistent with the scenario that most calcium was synthesized in inner layers and then partially escaped to the outermost layer. However, further investigation and theoretical analysis are needed to determine the true reasons of these results.

\section*{Acknowledgements}
	We thank Keiichi Maeda, Xiaofeng Wang and the anonymous referee for their insightful suggestions which help improve the	paper a lot. This research has made use of the CfA Supernova Archive, which is funded in part by the National Science Foundation through grant AST 0907903, the Berkeley/Lick Supernova Archives, which is funded in part by the US National Science Foundation, and the CSP Supernova Archive, which is supported by the World Premier International Research Center Initiative. Here we express our deep gratitude.

\section*{Data Availability}
The authors confirm that the data supporting the findings of this study are available within the article and its supplementary materials.

\clearpage
\onecolumn
\setlength{\tabcolsep}{2pt}
\begin{longtable}{ccccccccccccc}
    \caption{Measured Velocities and line strengths of Ca II H\&K}
    \label{Tab1}
	\\ \hline
	SN & Phase$^a$ & $V_{PHO}$~$^b$ & $W_{PHO}$~$^c$ & $V_{HVF}$~$^d$ & $W_{HVF}$~$^e$ & | & SN & Phase$^a$ & $V_{PHO}$~$^b$ & $W_{PHO}$~$^c$ & $V_{HVF}$~$^d$ & $W_{HVF}$~$^e$ \\
	~ &(days) & (km s$^{-1}$) & (\AA) & (km s$^{-1}$) & (\AA)
	& | & ~ &(days) & (km s$^{-1}$) & (\AA) & (km s$^{-1}$) & (\AA) \\
	\hline
1989B & -1 & 9867(75) & 40(2) & 17404(132) & 75(2) & | & 2006cj & -0.2 & 11776(322) & 72(5) & 19277(526) & 37(3) \\
1989M & -1 & 10437(86) & 25(1) & 17109(140) & 109(3) & | & 2006cj & 4.7 & 11577(281) & 64(4) & 18468(447) & 29(2) \\
1994D & -12.5 & 15000(288) & 36(2) & 25021(480) & 155(7) & | & 2006cz & -1 & 14179(340) & 64(4) & 20556(493) & 28(2) \\
1994D & -11.5 & 15000(113) & 36(1) & 24906(188) & 162(4) & | & 2006dm & -7.9 & 15000(603) & 96(9) & 25505(1024) & 71(7) \\
1994D & -9.5 & 12709(39) & 44(1) & 24368(76) & 129(2) & | & 2006dv & 0.4 & 11054(181) & 58(3) & 17508(287) & 24(1) \\
1994D & -8.5 & 11161(53) & 33(1) & 23205(109) & 126(2) & | & 2006dy & -11.8 & 14590(232) & 73(3) & 23825(378) & 91(4) \\
1994S & 0.7 & 10239(57) & 56(2) & 18295(102) & 54(2) & | & 2006ej & 4.8 & 11703(245) & 66(3) & 19139(401) & 20(1) \\
1994ae & -0.5 & 10396(248) & 60(4) & 17000(406) & 50(4) & | & 2006em & 3.7 & 11386(459) & 183(17) & 26611(1072) & 51(5) \\
1994ae & 0.5 & 10437(263) & 59(5) & 17000(429) & 55(4) & | & 2006et & 4.3 & 11337(239) & 51(3) & 18248(384) & 55(3) \\
1995D & 3.3 & 10224(47) & 52(1) & 17000(78) & 46(1) & | & 2006is & 4.6 & 12956(179) & 60(3) & 19927(274) & 49(2) \\
1995D & 4.3 & 10849(187) & 61(3) & 17000(294) & 29(2) & | & 2006kf & -8.3 & 12510(174) & 34(2) & 22015(305) & 129(4) \\
1996X & -0.1 & 11656(95) & 64(2) & 18561(150) & 12(1) & | & 2006kf & 0.7 & 10378(212) & 40(2) & 18114(371) & 51(3) \\
1996X & 0.9 & 11497(97) & 65(2) & 18314(154) & 12(1) & | & 2006oa & -0.1 & 9814(380) & 48(4) & 19579(760) & 61(6) \\
1998dk & -0.1 & 11363(90) & 23(1) & 18912(148) & 116(3) & | & 2006or & 0.3 & 11530(86) & 53(2) & 19385(144) & 74(2) \\
1998es & -9.4 & 9308(154) & 23(1) & 19695(326) & 64(3) & | & 2006sr & -1 & 11208(201) & 34(2) & 18663(335) & 72(4) \\
1998es & -0.7 & 10298(207) & 42(2) & 18644(375) & 47(2) & | & 2006sr & 1 & 11560(219) & 51(2) & 18782(356) & 54(2) \\
1999aa & -11.4 & 11159(211) & 19(1) & 21301(402) & 50(3) & | & 2006sr & 3 & 12492(309) & 68(4) & 19344(477) & 30(2) \\
1999aa & -10 & 10837(201) & 20(1) & 21075(391) & 45(2) & | & 2007A & 0.6 & 10845(58) & 61(2) & 17537(93) & 31(1) \\
1999aa & -9.1 & 10535(226) & 21(1) & 21024(451) & 44(3) & | & 2007F & -9.7 & 12391(78) & 39(1) & 23276(146) & 65(2) \\
1999aa & -8.1 & 9755(144) & 25(1) & 20616(303) & 42(2) & | & 2007F & -8.9 & 10932(412) & 30(3) & 22031(831) & 93(9) \\
1999aa & -0.4 & 11158(132) & 56(2) & 19697(233) & 39(1) & | & 2007F & -7.9 & 10937(241) & 32(2) & 21646(477) & 72(4) \\
1999cp & -12 & 10598(125) & 27(2) & 18343(216) & 90(3) & | & 2007F & -0.9 & 10963(193) & 57(3) & 18744(329) & 45(2) \\
1999cp & 4.8 & 10418(150) & 61(3) & 17000(243) & 26(1) & | & 2007F & 3 & 10865(103) & 61(2) & 17485(165) & 42(2) \\
1999dk & -7.7 & 14302(103) & 35(1) & 23296(167) & 150(3) & | & 2007af & -11.1 & 14681(118) & 73(2) & 23275(187) & 85(2) \\
1999ee & -9.4 & 16000(65) & 37(1) & 23719(95) & 84(2) & | & 2007bc & 0.4 & 10454(99) & 64(2) & 17212(162) & 43(1) \\
1999ee & -0.4 & 15508(97) & 76(1) & 21151(132) & 32(1) & | & 2007bd & -8.1 & 11471(257) & 22(2) & 19871(446) & 114(7) \\
2000cx & 4.2 & 12570(616) & 22(2) & 19845(972) & 29(3) & | & 2007bm & -8.2 & 10647(134) & 40(2) & 18653(235) & 60(2) \\
2000dk & 0.4 & 10657(101) & 39(1) & 17301(164) & 93(3) & | & 2007ca & 4.3 & 10676(87) & 61(2) & 17399(141) & 44(2) \\
2000dp & 1 & 10842(149) & 62(2) & 17758(243) & 63(2) & | & 2007co & -0.2 & 11024(276) & 38(3) & 19005(477) & 108(7) \\
2000dx & -9 & 15000(680) & 54(6) & 26126(1183) & 142(15) & | & 2007co & 0.8 & 11250(174) & 48(2) & 19087(295) & 96(4) \\
2000ey & 0 & 13229(428) & 68(5) & 19713(637) & 16(1) & | & 2007co & 3.7 & 10378(262) & 37(3) & 17933(452) & 97(7) \\
2000fa & -10.7 & 15000(325) & 37(2) & 25659(556) & 144(8) & | & 2007co & 4.7 & 12492(416) & 76(7) & 18938(630) & 51(5) \\
2000fa & -9.1 & 15000(152) & 51(2) & 23947(243) & 114(4) & | & 2007fr & -0.2 & 13113(1139) & 93(18) & 28000(2432) & 5(1) \\
2001ay & 3.4 & 11696(120) & 32(1) & 20155(206) & 64(2) & | & 2007gi & -7.8 & 15000(1052) & 66(9) & 27216(1908) & 154(22) \\
2001ba & 3.8 & 9996(95) & 60(2) & 17000(160) & 66(2) & | & 2007gi & -0.8 & 8000(356) & 19(2) & 19950(887) & 132(12) \\
2001br & 3.1 & 14412(564) & 87(8) & 21015(822) & 8(1) & | & 2007gk & -0.8 & 12159(82) & 35(2) & 19810(133) & 84(2) \\
2001cp & 0.6 & 10640(105) & 53(2) & 17627(174) & 56(2) & | & 2007hu & 3.1 & 11223(129) & 29(1) & 19507(224) & 106(3) \\
2001da & -1 & 13193(134) & 64(2) & 21641(220) & 100(3) & | & 2007kk & -0.8 & 11949(242) & 39(2) & 19763(401) & 97(5) \\
2001da & -0.5 & 13366(242) & 77(4) & 21447(389) & 86(4) & | & 2007kk & 3.3 & 13628(373) & 81(6) & 20681(566) & 52(4) \\
2002av & -1 & 10290(285) & 61(5) & 17000(469) & 72(6) & | & 2007kk & 4.3 & 11337(361) & 53(5) & 18804(600) & 89(7) \\
2002cd & 0.8 & 13940(717) & 54(7) & 20939(1076) & 77(9) & | & 2007le & -10.4 & 15000(61) & 24(1) & 24704(118) & 143(2) \\
2002ck & 4 & 10971(183) & 66(3) & 17000(282) & 28(2) & | & 2007le & -9.7 & 15000(702) & 52(6) & 27264(1276) & 158(16) \\
2002de & -0.1 & 11021(133) & 43(2) & 18999(229) & 69(3) & | & 2007le & -9.5 & 15000(564) & 50(4) & 27193(1023) & 157(12) \\
2002dj & -10.6 & 18266(279) & 76(4) & 28000(428) & 118(4) & | & 2007le & 3.3 & 10983(145) & 37(2) & 18540(244) & 103(4) \\
2002dj & -9.6 & 16441(423) & 40(3) & 25578(658) & 158(10) & | & 2007le & 4.3 & 10812(99) & 33(1) & 18412(168) & 111(3) \\
2002dj & -7.6 & 14090(88) & 27(1) & 22722(143) & 155(3) & | & 2007nq & 3.8 & 11854(311) & 46(3) & 19009(498) & 63(4) \\
2002eb & 1 & 10214(92) & 44(1) & 18243(163) & 62(2) & | & 2007ux & 0 & 12592(1025) & 83(15) & 26411(2150) & 14(3) \\
2002er & -7.9 & 15000(248) & 90(3) & 24142(400) & 61(3) & | & 2008Y & 4.4 & 10174(370) & 26(3) & 18711(680) & 138(12) \\
2002er & 4.1 & 12398(155) & 83(3) & 18356(229) & 23(1) & | & 2008ar & -8.7 & 17677(242) & 110(4) & 26143(359) & 63(3) \\
2002eu & 0.7 & 9323(126) & 24(2) & 17856(242) & 91(3) & | & 2008ar & -7.7 & 17222(434) & 111(7) & 25428(641) & 60(4) \\
2002fb & 0.9 & 11572(1631) & 119(35) & 28000(3947) & 54(16) & | & 2008ar & -0.7 & 12212(354) & 66(4) & 21251(615) & 93(6) \\
2002ha & -0.4 & 10550(55) & 51(2) & 17507(91) & 62(2) & | & 2008ar & 3.3 & 10241(252) & 40(2) & 18857(465) & 112(7) \\
2002hd & 4.9 & 11119(99) & 69(2) & 17435(154) & 10(1) & | & 2008ar & 4.3 & 10669(234) & 53(3) & 19177(421) & 106(6) \\
2002he & 0.2 & 13618(190) & 84(3) & 20114(281) & 23(1) & | & 2008bc & -9.7 & 14613(256) & 60(3) & 26006(455) & 151(7) \\
2002he & 3.2 & 13317(174) & 82(3) & 19780(258) & 17(1) & | & 2008bc & 4.3 & 10064(152) & 39(2) & 17000(256) & 83(3) \\
2003U & -8.8 & 12594(72) & 39(1) & 20953(120) & 119(2) & | & 2008bf & 0.7 & 11449(91) & 59(2) & 20137(159) & 64(2) \\
2003du & -9.1 & 11006(207) & 52(3) & 20815(390) & 90(4) & | & 2008bf & 3.7 & 10424(133) & 36(2) & 17978(229) & 76(3) \\
2003du & -0.1 & 9922(154) & 52(3) & 18174(281) & 74(4) & | & 2008bf & 4.7 & 10612(116) & 45(2) & 18263(200) & 65(2) \\
2003fa & -9.2 & 12420(196) & 30(2) & 20985(332) & 40(2) & | & 2008ec & -0.1 & 10700(118) & 63(2) & 17724(195) & 52(2) \\
2003kf & -8.8 & 12198(80) & 52(2) & 22289(146) & 90(2) & | & 2008ei & 0.6 & 12113(465) & 32(3) & 20569(789) & 140(13) \\
2004E & 4.3 & 11156(315) & 57(4) & 18472(523) & 14(2) & | & 2008hj & -10.7 & 13630(184) & 53(2) & 24169(326) & 119(4) \\
2004S & -0.9 & 15000(214) & 95(3) & 21653(309) & 53(3) & | & 2008hv & -11.3 & 15000(189) & 39(2) & 23827(300) & 115(4) \\
2004dt & -10.8 & 15000(250) & 7(1) & 23284(124) & 89(4) & | & 2009aa & -7.7 & 10532(140) & 59(2) & 18324(243) & 15(1) \\
2004dt & -9.8 & 15000(460) & 12(1) & 23022(127) & 89(7) & | & 2009ig & -9.5 & 15000(215) & 41(1) & 26889(386) & 154(6) \\
2004dt & -7.8 & 11232(166) & 8(1) & 21527(95) & 86(3) & | & 2011by & -12.4 & 12506(187) & 35(2) & 20521(307) & 86(4) \\
2004ef & -0.4 & 10460(175) & 21(2) & 20160(337) & 132(6) & | & 2011fe & -12 & 12998(31) & 49(1) & 21010(50) & 92(1) \\
2004ef & 0.6 & 11514(267) & 42(3) & 20470(475) & 103(6) & | & 2011fe & -11 & 12133(29) & 46(1) & 20095(48) & 84(1) \\
2004ey & -8.2 & 12582(66) & 53(1) & 22421(117) & 92(1) & | & 2011fe & -10 & 12182(29) & 56(1) & 20111(47) & 67(1) \\
2004fg & -1 & 10458(71) & 48(2) & 17457(119) & 40(2) & | & 2011fe & -9 & 11484(33) & 54(1) & 19348(55) & 57(1) \\
2005cf & -10.7 & 15000(276) & 58(3) & 24515(451) & 108(5) & | & 2011fe & -8 & 10630(37) & 43(1) & 18586(64) & 58(1) \\
2005cf & -9.7 & 15000(179) & 69(3) & 24198(288) & 101(3) & | & 2011fe & -1 & 10615(30) & 63(1) & 17806(50) & 37(1) \\
2005cf & 3.3 & 10030(134) & 47(2) & 18679(249) & 93(3) & | & 2011fe & 0 & 10314(30) & 58(1) & 17352(49) & 40(1) \\
2005cf & 4.3 & 10148(80) & 57(1) & 18984(102) & 91(2) & | & 2011fe & 1 & 10342(33) & 61(1) & 17247(54) & 39(1) \\
2005de & -0.9 & 9544(157) & 25(2) & 17000(279) & 122(6) & | & 2011fe & 3 & 10626(82) & 62(1) & 17000(131) & 27(1) \\
2005df & -10.8 & 14656(45) & 42(1) & 24557(74) & 142(2) & | & 2011fe & 4 & 10747(139) & 63(2) & 17000(220) & 23(1) \\
2005df & -7.8 & 15000(42) & 80(1) & 24351(68) & 101(2) & | & 2011iv & 5 & 12467(115) & 74(2) & 18918(174) & 16(1) \\
2005ej & 3.1 & 10861(215) & 73(4) & 17507(346) & 28(2) & | & 2011jh & 3.4 & 11575(215) & 20(2) & 19336(359) & 129(6) \\
2005eq & -8.5 & 11240(289) & 26(2) & 20307(522) & 49(3) & | & 2012bh & 0.3 & 10542(144) & 61(2) & 17000(232) & 41(2) \\
2005eq & 0.4 & 11108(224) & 52(3) & 19240(387) & 44(3) & | & 2012cg & -8.3 & 13923(109) & 57(2) & 22082(172) & 57(2) \\
2005na & -0.6 & 10713(175) & 61(3) & 18354(299) & 20(1) & | & 2012fr & 0 & 13855(349) & 50(3) & 23013(579) & 88(5) \\
2005na & 0.5 & 10824(309) & 69(5) & 18625(532) & 20(2) & | & 2012fr & 3 & 13050(258) & 51(3) & 22540(446) & 100(5) \\
2006S & -0.7 & 11801(246) & 60(4) & 19881(414) & 57(3) & | & 2012hm & 5 & 10612(226) & 21(2) & 20734(440) & 151(8) \\
2006S & 0.3 & 11694(173) & 60(2) & 19694(290) & 59(2) & | & 2012hq & 0 & 14995(313) & 25(1) & 22550(471) & 45(2) \\
2006S & 3.3 & 11490(358) & 58(5) & 19339(603) & 58(5) & | & 2013dy & -10.5 & 12466(131) & 42(2) & 21484(226) & 49(2) \\
2006X & 0.9 & 12967(291) & 46(3) & 20971(470) & 111(6) & | & 2013dy & -9.5 & 12313(192) & 44(2) & 21433(335) & 37(1) \\
2006ax & -10.4 & 14904(149) & 87(3) & 22068(221) & 44(2) & | & 2013dy & -7.5 & 11654(171) & 45(2) & 21140(309) & 36(2) \\
2006az & -1 & 10326(281) & 61(4) & 17710(482) & 22(2) & | & 2013gs & -7.5 & 15000(152) & 53(2) & 24911(252) & 126(4) \\
2006bu & 3.9 & 10301(216) & 46(3) & 17000(357) & 61(3) & | & 2015bq & -12.2 & 14600(258) & 44(3) & 24448(432) & 27(1) \\
2006cf & 3.9 & 10193(172) & 60(3) & 17059(288) & 47(3) & | & 2015bq & -10.2 & 14568(180) & 39(1) & 22418(277) & 28(1) \\
2006cf & 4.9 & 10370(254) & 60(4) & 17107(419) & 38(3) & | & 2018oh & -8.6 & 8695(123) & 27(2) & 17998(254) & 91(4) \\
	\hline
	\multicolumn{13}{l}{Data sources and photometric parameters can be found in \citet{Zhao16,Zhao21}. }\\
	\multicolumn{13}{l}{$^a$ Days since $B-$band maximum light;}\\
	\multicolumn{13}{l}{\footnotesize$^b$ Velocity of the PHO component of Ca II H\&K; }\\
	\multicolumn{13}{l}{\footnotesize$^c$ Pseudo-equivalent width of the PHO component of Ca II H\&K; }\\
	\multicolumn{13}{l}{\footnotesize$^d$ Velocity of the HVF component of Ca II H\&K;}\\
	\multicolumn{13}{l}{\footnotesize$^e$ Pseudo-equivalent width of the HVF component of Ca II H\&K;}\\
\end{longtable}

\end{document}